\documentclass[conference]{IEEEtran}
\IEEEoverridecommandlockouts
\usepackage{cite}
\usepackage{amsmath,amssymb,amsfonts}
\usepackage{graphicx}
\usepackage{textcomp}
\usepackage{xcolor}
\usepackage{booktabs}
\usepackage{url}
\usepackage{balance}
\usepackage{dblfloatfix}   

\def\BibTeX{{\rm B\kern-.05em{\sc i\kern-.025em b}\kern-.08em T\kern-.1667em\lower.7ex\hbox{E}\kern-.125emX}}

\begin{document}

\title{X-KGRank: A Knowledge Graph RAG Framework for Explainable Recommendations via Pattern Mining and LLM Re-Ranking}

\author{%
\IEEEauthorblockN{Meenakshi Rajpurohit}
\IEEEauthorblockA{Dept.\ of Computer Engineering\\
San Jose State University\\
San Jose, CA, USA\\
meenakshi.rajpurohit@sjsu.edu}
\and
\IEEEauthorblockN{Jainish Patel}
\IEEEauthorblockA{Dept.\ of Computer Engineering\\
San Jose State University\\
San Jose, CA, USA\\
jainish.patel@sjsu.edu}

}

\maketitle

\begin{abstract}
Modern recommender systems produce predictions that users cannot interrogate.
The two dominant improvements, collaborative filtering and LLM-based reasoning,
each fall short: collaborative filtering captures behavioural signals but offers
no reasoning, while large language models (LLMs) generate fluent explanations but
hallucinate and are poorly grounded in a user's history. We present X-KGRank, a
knowledge-graph retrieval-augmented framework that unifies structural collaborative
filtering with LLM-based explanation. From the MovieLens-1M dataset (6{,}040 users,
3{,}704 items, 988{,}129 interactions) we construct a heterogeneous knowledge graph
of 9{,}762 nodes and 999{,}264 edges spanning three relation types---RATED,
HAS\_GENRE, and CO\_RATED---persisted in Neo4j. We train a LightGCN ranker with
content-aware SBERT initialisation and a rating-weighted BPR objective, and apply a
popularity-selective routing strategy that grounds long-tail items (1{,}855 of
3{,}704) in knowledge-graph paths while serving popular items from pretrained
knowledge, reducing KG-augmented generations by roughly 50\%. On the MovieLens-1M
test set under a 99-sample protocol, X-KGRank achieves NDCG@10 $=0.2956$ and
Recall@10 $=0.5371$, improving over a strong popularity baseline by 17.1\% on both
metrics, by 15.6\% on NDCG@20 ($0.3449$ vs.\ $0.2983$), and by 14.6\% on MRR
($0.2435$ vs.\ $0.2124$). Across three LLM backbones evaluated on 16 cases, a
1.5-billion-parameter model (Qwen2.5-1.5B) matches a 7-billion-parameter model
(Mistral-7B) on heuristic explanation quality ($0.97$ vs.\ $0.94$), yet qualitative
analysis shows the smaller model is more prone to factual fabrication.
\end{abstract}


\section{Introduction}

\begin{figure*}[!tb]
\centering
\includegraphics[width=\textwidth]{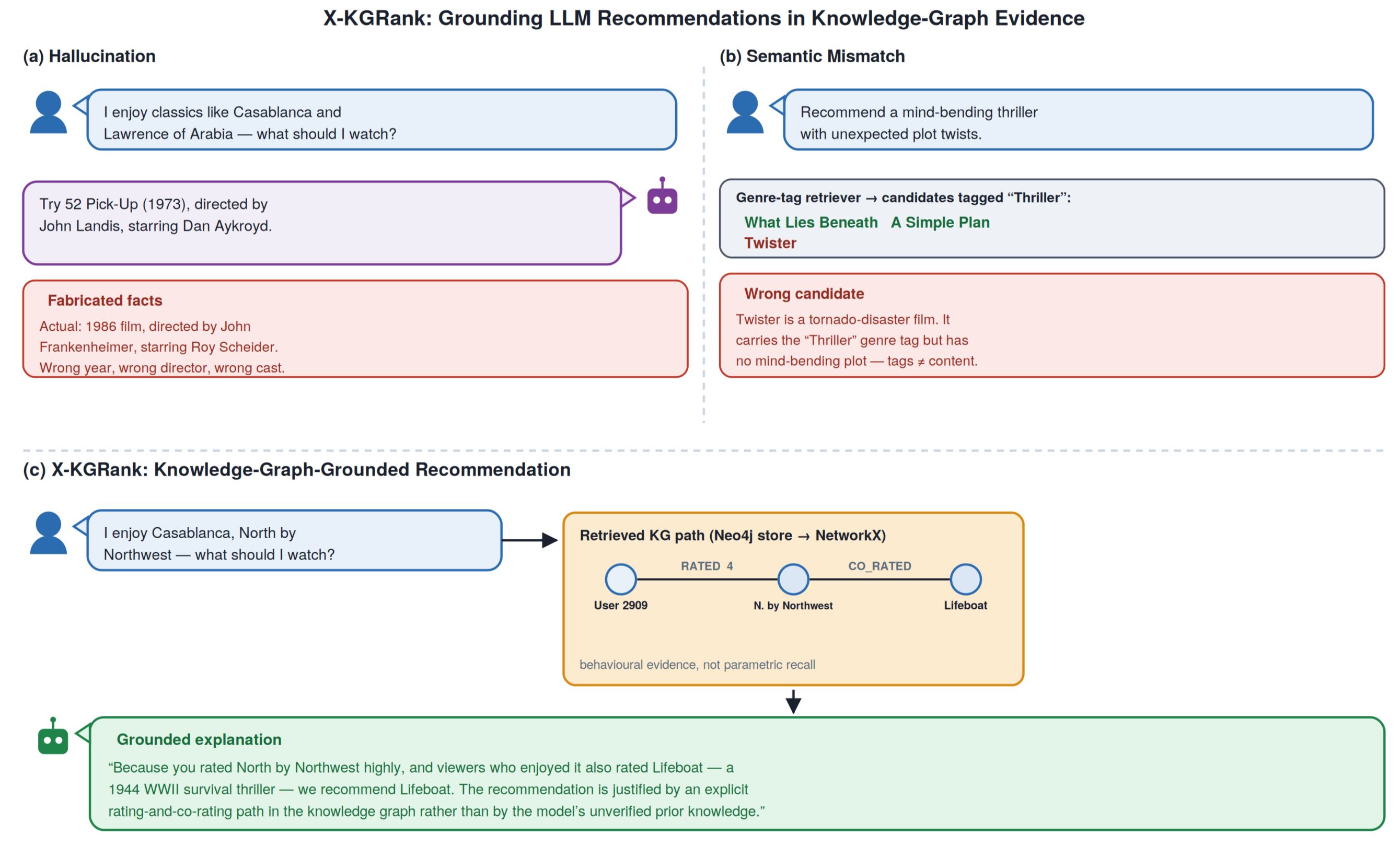}
\caption{Two failure modes of LLM-based recommendation and X-KGRank's
knowledge-graph-grounded solution. (a)~The LLM fabricates a film's director, cast, and
year. (b)~Genre-tag retrieval returns \emph{Twister}, a disaster film tagged
``Thriller,'' for a mind-bending-thriller query. (c)~X-KGRank retrieves an explicit
rating-and-co-rating path and conditions its explanation on that behavioural evidence.
All examples are drawn from our experiments.}
\label{fig:motivation}
\end{figure*}

When a user asks a recommender for ``a mind-bending thriller with unexpected plot
twists,'' what happens? A content-based system that relies only on genre tags labelled
``Thriller'' may retrieve both a genuine psychological thriller such as
\emph{What Lies Beneath} and an unrelated disaster film such as \emph{Twister}, because
both carry the same tag. A collaborative-filtering model ranks items by behavioural
similarity but offers no user-specific explanation of why a movie was recommended.
An LLM-based recommender can generate fluent explanations, but it hallucinates and is
factually unreliable: it may miss release years, invent cast members, or attribute
films to the wrong directors. The results have limited interpretability and give no
clear signal of why they should be trusted. These failure modes arise directly from
our own experimental analysis, illustrated in Fig.~\ref{fig:motivation}.

Existing frameworks each address part of this problem but each leaves a critical gap.
Graph-based collaborative-filtering methods such as LightGCN~\cite{lightgcn},
KGAT~\cite{kgat}, and GraphSAGE~\cite{graphsage} provide strong behavioural ranking
signals, yet their effectiveness degrades as interaction sparsity increases. LLM-based
recommenders such as P5~\cite{p5} offer good natural-language reasoning, but they
hallucinate and often do not ground recommendations in a user's specific interaction
history. Knowledge-graph-based recommenders incorporate structural information into
ranking, but they typically use the graph as model input rather than as evidence for
explanation generation.

A key observation behind our approach is that not every item requires the same level of
expensive grounding to produce a trustworthy explanation. For popular items, an LLM
often already possesses the relevant pretrained knowledge; for long-tail (less popular)
items, that pretrained knowledge is less reliable, and knowledge-graph evidence becomes
more valuable. A popularity-selective routing strategy---open LLM prompts for popular
candidates and KG-grounded prompts for long-tail candidates---focuses retrieval where it
is most useful. We adopt and extend this design, inspired by K-RagRec~\cite{kragrec},
into a complete recommendation framework.

We present X-KGRank, a six-stage knowledge-graph retrieval-augmented framework for
explainable LLM recommendation. X-KGRank constructs a heterogeneous knowledge graph in
Neo4j, trains LightGCN embeddings initialised from SBERT-derived item representations,
mines structural and community-level graph features using node2vec and modularity-based
clustering, retrieves candidate-specific knowledge-graph paths at inference time, and
routes each candidate through one of two LLM explanation prompts according to its
popularity bucket. On MovieLens-1M, X-KGRank improves over a strong baseline by 17\% on
NDCG@10 while also producing user-specific natural-language explanations grounded in
graph evidence. Our work makes three main contributions: (i)~an end-to-end framework for
knowledge-graph retrieval-augmented recommendation; (ii)~a popularity-selective routing
strategy; and (iii)~an explanation-quality analysis across the Flan-T5-large,
Qwen2.5-1.5B, and Mistral-7B backbones.

\section{Related Work}

The literature relevant to X-KGRank spans three intersecting lines of research: graph
neural networks for structural collaborative filtering, LLM-based recommendation, and
retrieval-augmented generation over knowledge graphs.

\subsection{Graph Neural Networks for Recommendation}
Graph neural networks have become a dominant framework for collaborative filtering,
treating user--item interactions as a graph and propagating embeddings through multi-hop
neighbourhoods. He et al.~\cite{lightgcn} introduced LightGCN, which simplifies standard
graph convolution by removing non-linear activations and feature transformations,
retaining only neighbourhood aggregation; this achieves state-of-the-art performance
while reducing model complexity. We adopt LightGCN as our ranker and add LLM-based
explanations that operate on retrieved KG paths.

\subsection{LLMs for Recommendation}
Recent work explores LLMs as recommenders. Geng et al.~\cite{p5} proposed P5, unifying
five recommendation tasks---sequential recommendation, rating prediction, explanation
generation, review-related tasks, and direct recommendation---under a single
text-to-text framework. While flexible, the pure-LLM approach has two weaknesses: it
hallucinates, and it lacks grounding in user-specific behavioural signals. Our work
addresses these weaknesses by retaining a structural model for ranking and confining the
LLM to an explanatory role over retrieved KG evidence.

\subsection{Retrieval-Augmented Generation over Knowledge Graphs}
Retrieval-augmented generation (RAG) has emerged as a remedy for both the sparsity limits
of pure structural methods and the hallucination problems of pure-LLM methods.
He et al.~\cite{gretriever} proposed G-Retriever, which retrieves task-relevant subgraphs
and conditions an LLM on them for textual graph question answering, showing an advantage
over dense vector retrieval. Most directly related, Wang et al.~\cite{kragrec} introduced
K-RagRec, a knowledge-graph retrieval-augmented framework for LLM-based recommendation
that applies expensive subgraph retrieval selectively to cold-start items while serving
popular items through cheaper structural ranking. X-KGRank builds on the K-RagRec design
along two axes. First, we provide an open implementation on MovieLens-1M, including the
Neo4j graph database, the LightGCN ranker, the popularity-selective routing strategy, and
the MLP projection layer. Second, we evaluate the explanation layer across three LLMs
(Flan-T5-large, Qwen2.5-1.5B, Mistral-7B), surfacing a previously unreported trade-off
between explanation quality and factual reliability in small models.

\section{Preliminaries and Problem Formulation}

\subsection{Knowledge Graph and Notation}
Let $\mathcal{U}=\{u_1,\dots,u_n\}$ denote the set of users, $\mathcal{I}=\{i_1,\dots,i_m\}$
the set of movies, and $\mathcal{C}$ the set of content categories (genres). We model
their relationships as a heterogeneous knowledge graph
\begin{equation}
\mathcal{G}=(\mathcal{V},\mathcal{E},\tau),
\qquad
\mathcal{V}=\mathcal{U}\cup\mathcal{I}\cup\mathcal{C},
\end{equation}
where each edge is assigned a relation type by the function
$\tau:\mathcal{E}\rightarrow\mathcal{R}$ with relation set
\begin{equation}
\mathcal{R}=\{\,\mathrm{RATED},\ \mathrm{HAS\_GENRE},\ \mathrm{CO\_RATED}\,\}.
\end{equation}
A RATED edge $(u,i)$ carries an integer rating from $1$ to $5$; a HAS\_GENRE edge $(i,c)$
links an item to a genre; and a CO\_RATED edge $(i,j)$ carries a weight $w_{ij}$ equal to
the number of users who rated both items.

\section{Methodology}

\begin{figure*}[!tb]
\centering
\includegraphics[width=\textwidth]{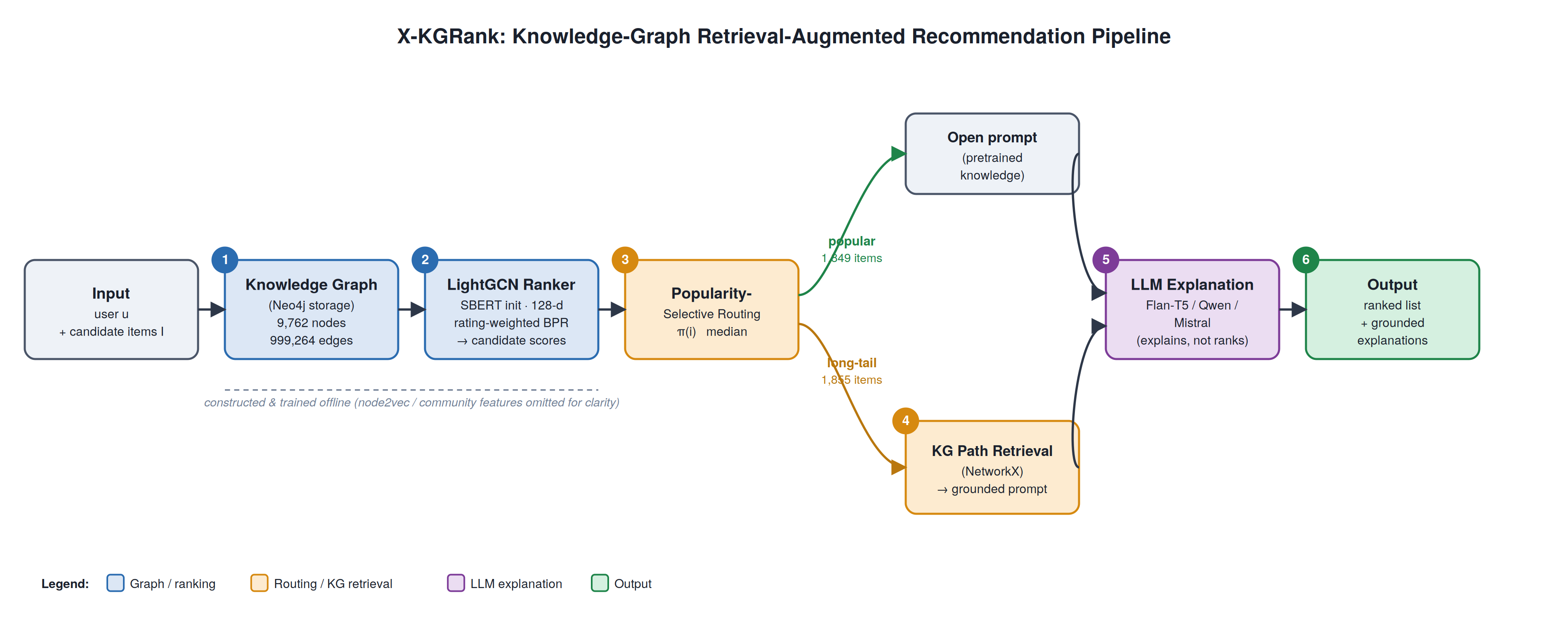}
\caption{High-level overview of the X-KGRank pipeline. A user and a candidate item set
enter a LightGCN ranker trained over a knowledge graph stored in Neo4j. Each candidate is
routed by popularity; the LLM generates explanations while the final ranking is determined
by LightGCN scores.}
\label{fig:pipeline}
\end{figure*}

\begin{figure*}[!tb]
\centering
\includegraphics[width=\textwidth]{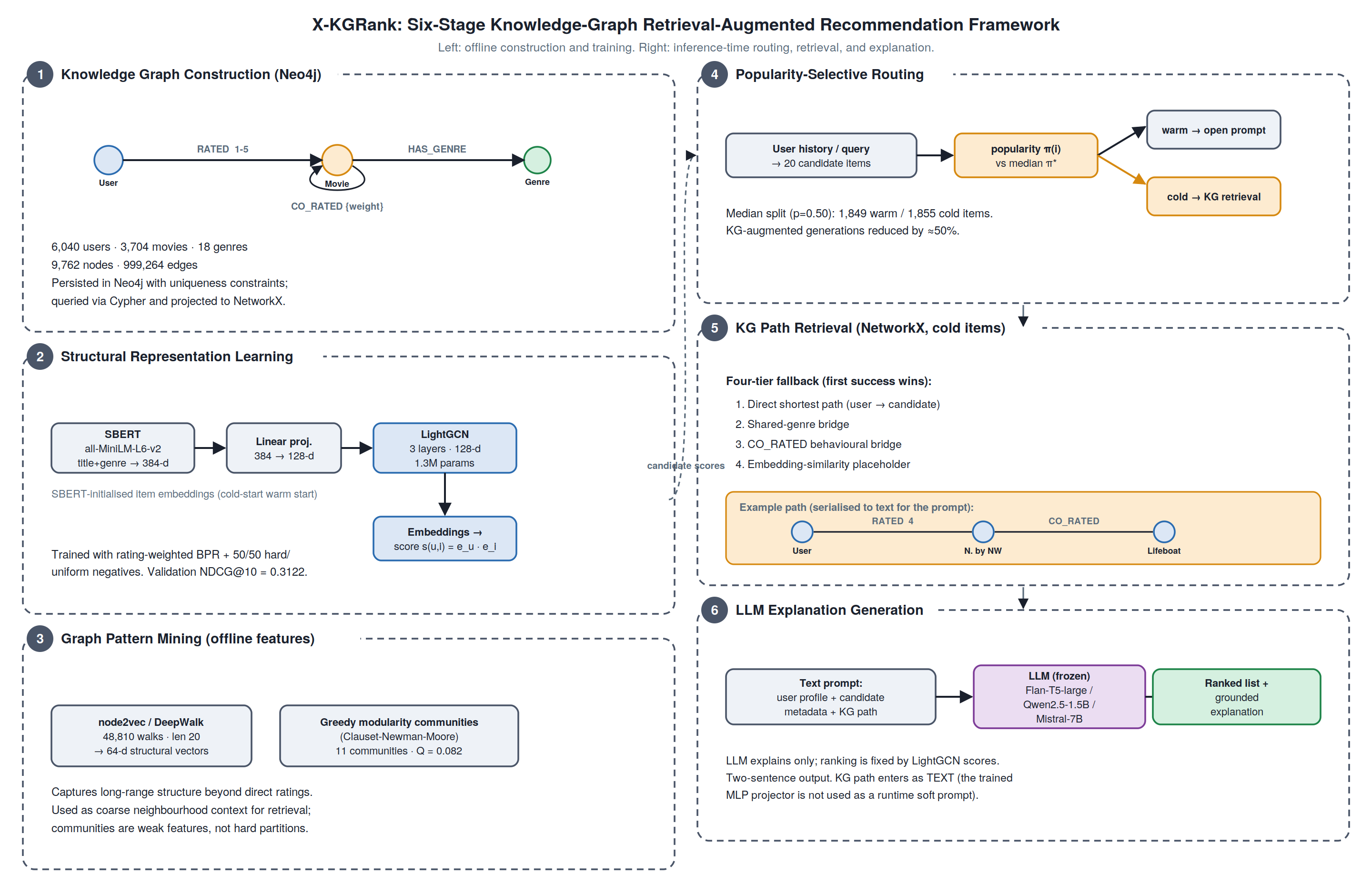}
\caption{Detailed X-KGRank architecture. Offline (left): (1)~a heterogeneous knowledge
graph is stored in Neo4j; (2)~a LightGCN ranker is trained with SBERT-initialised
embeddings; (3)~node2vec and community features are mined. At inference (right):
(4)~candidates are routed by popularity; (5)~cold items trigger a four-tier KG path
retrieval over an in-memory NetworkX projection; (6)~the retrieved path is serialised into
a text prompt for explanation, while ranking stays fixed by LightGCN scores.}
\label{fig:arch}
\end{figure*}

In this section we introduce the key concepts used to build the system and then describe
each component of the framework in detail. Fig.~\ref{fig:pipeline} gives a high-level
overview of the complete pipeline, and Fig.~\ref{fig:arch} expands each stage with
implementation detail.

\subsection{Knowledge Graph Construction}
\begin{figure}[!htbp]
\centering
\includegraphics[width=\columnwidth]{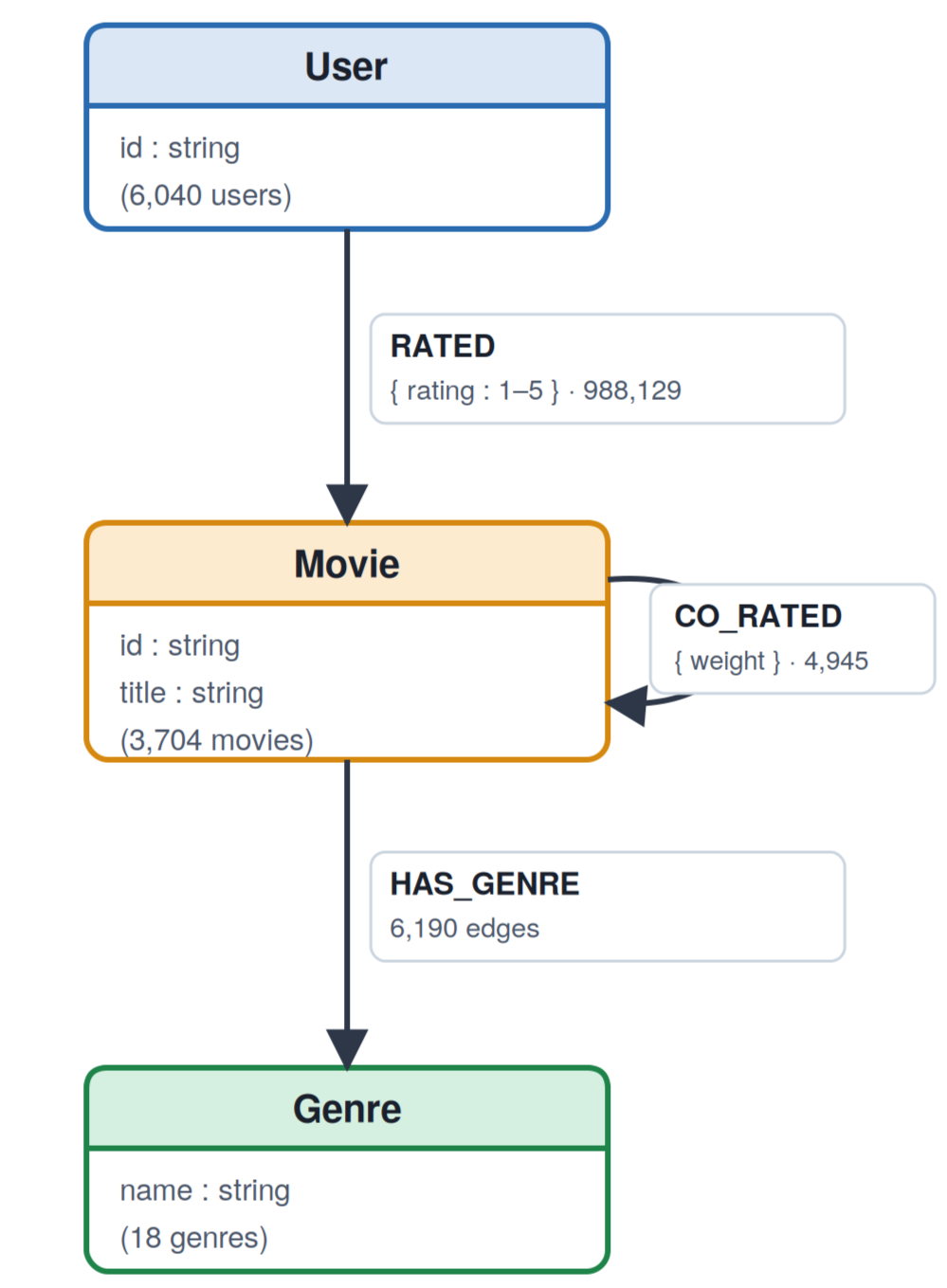}
\caption{Schema of the X-KGRank knowledge graph. Three node types (User, Movie, Genre) are
connected by three relation edges: RATED (User\,$\rightarrow$\,Movie),
HAS\_GENRE (Movie\,$\rightarrow$\,Genre), and CO\_RATED (Movie\,$\rightarrow$\,Movie).}
\label{fig:schema}
\end{figure}

We construct a knowledge graph $\mathcal{G}=(\mathcal{V},\mathcal{E},\tau)$ whose vertex
set $\mathcal{V}=\mathcal{U}\cup\mathcal{I}\cup\mathcal{C}$ comprises users, items
(movies), and content nodes (genres). Edges are of three types: RATED, HAS\_GENRE, and
CO\_RATED. Fig.~\ref{fig:schema} shows the resulting schema and the three relation types.

\begin{figure}[!htbp]
\centering
\includegraphics[width=\columnwidth]{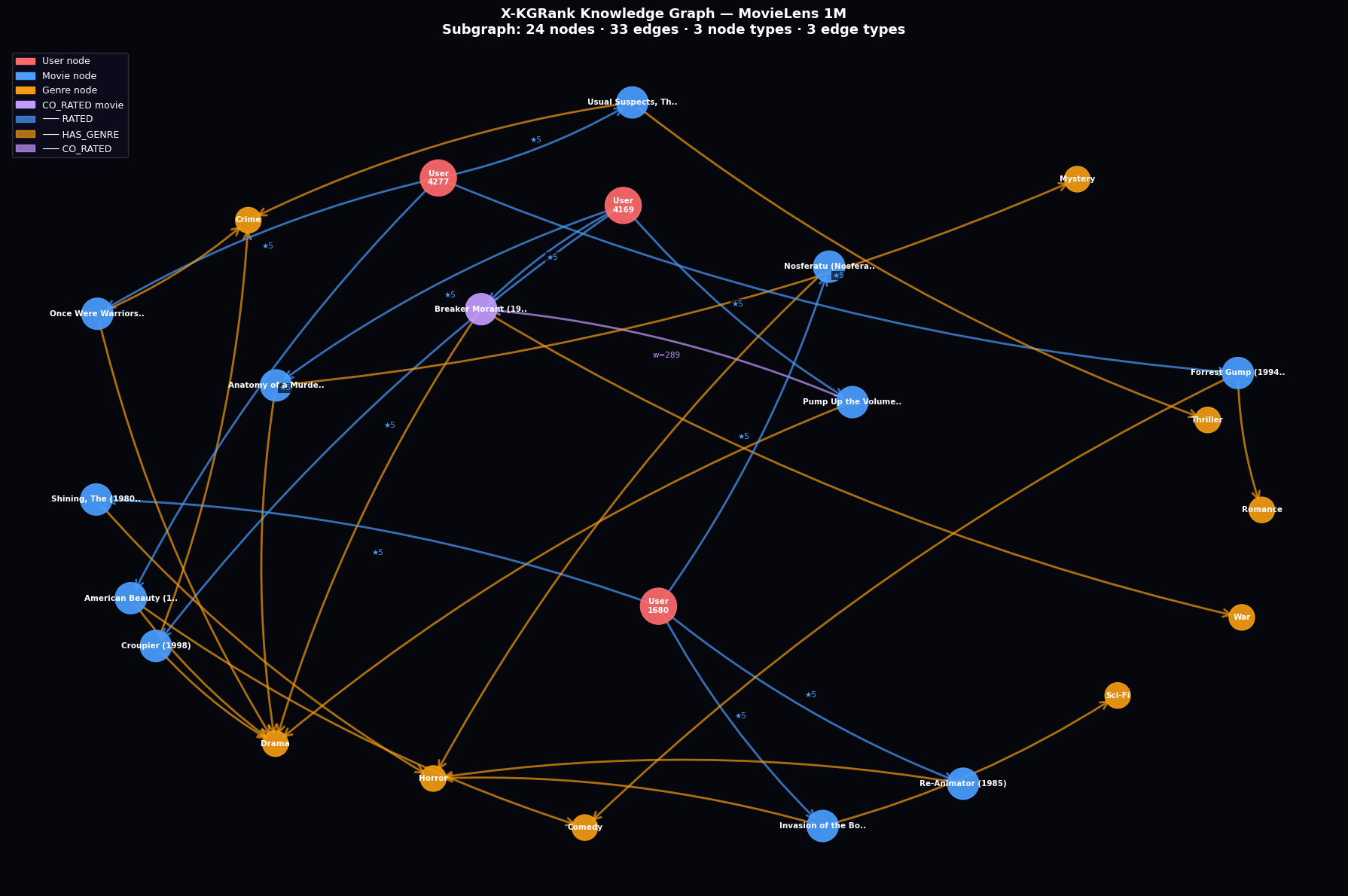}
\caption{A 24-node subgraph of the X-KGRank knowledge graph showing User, Movie, and Genre
nodes connected by RATED, HAS\_GENRE, and CO\_RATED edges.}
\label{fig:subgraph}
\end{figure}

The RATED relation carries ratings in the range $1$--$5$. The HAS\_GENRE edge connects an
item to its genre nodes. The CO\_RATED edge encodes behavioural co-occurrence and lets the
LLM generate the crucial ``users who rated $A$ also rated $B$'' explanation; without
CO\_RATED edges, the only available path between two movies would pass through user nodes,
which carry no semantic context for detailed explanations. For example, in MovieLens-1M,
User~1007 rated \emph{Raiders of the Lost Ark} (1981) five stars, and the knowledge graph
links this behaviour to related items, helping surface user-relevant relationships.

By converting the MovieLens-1M tables into a knowledge graph, Neo4j provides meaning and
context that plain lists and tables do not, allowing us to uncover connections useful for
recommendation. The full knowledge graph contains 9{,}762 nodes (6{,}040 users, 3{,}704
movies, and 18 genres) and 999{,}264 edges: 988{,}129 RATED, 6{,}190 HAS\_GENRE, and
4{,}945 CO\_RATED. This structured context provides the LLM with more relevant grounding
than semantic search alone.

\subsection{Structural Representation Learning}
LightGCN propagates information over a graph of users and items connected by rating edges.
It removes feature transformation and non-linear activation, keeping only neighbourhood
aggregation, which is well suited to user--item interactions. Our LightGCN uses $3$
propagation layers and an embedding dimension of $128$, yielding a model with
$1{,}296{,}384$ parameters.

Sentence-BERT (SBERT) turns text into vector embeddings that capture semantic meaning;
instead of starting from random weights, the model begins with sentence-level semantic
similarity. Each movie is represented using its title and genre, encoded with the
\texttt{all-MiniLM-L6-v2} SBERT model into a $384$-dimensional semantic vector, so movies
with similar titles and genres receive nearby representations. Because the recommender uses
a $128$-dimensional embedding space, a linear projection layer maps each $384$-dimensional
SBERT vector into the LightGCN dimension. Rather than initialising item embeddings randomly,
we therefore start from embeddings that already encode semantic information about each movie;
during training these embeddings are updated by user--item interactions, so the model learns
both user-preference patterns and movie content. This design directly targets the
cold-start problem: items with few ratings would otherwise behave as noise after training,
and SBERT initialisation gives such items a meaningful starting representation. The item
score for a user is the inner product of their learned embeddings,
\begin{equation}
s(u,i)=\mathbf{e}_u^{\top}\mathbf{e}_i,
\end{equation}
and the ranker is trained with a rating-weighted BPR objective using a 50/50 mix of hard and
uniform negatives. The validation NDCG@10 of the trained ranker is $0.3122$.

\subsection{Graph Pattern Mining}
LightGCN learns from direct user--movie interactions but does not capture long-range graph
patterns. We use two graph-mining techniques: node2vec/DeepWalk embeddings and greedy
modularity-based community detection.

node2vec performs random walks through the graph, traversing edges in sequence (e.g.,
user\,$\rightarrow$\,movie\,$\rightarrow$\,user). From $48{,}810$ random walks of length
$20$, a skip-gram model with negative sampling produces $64$-dimensional vectors for all
$9{,}762$ nodes. Movies with similar graph structure receive similar embeddings even when
no single user rated both, giving the system structural knowledge beyond direct ratings.

\begin{figure*}[!tb]
\centering
\includegraphics[width=\textwidth]{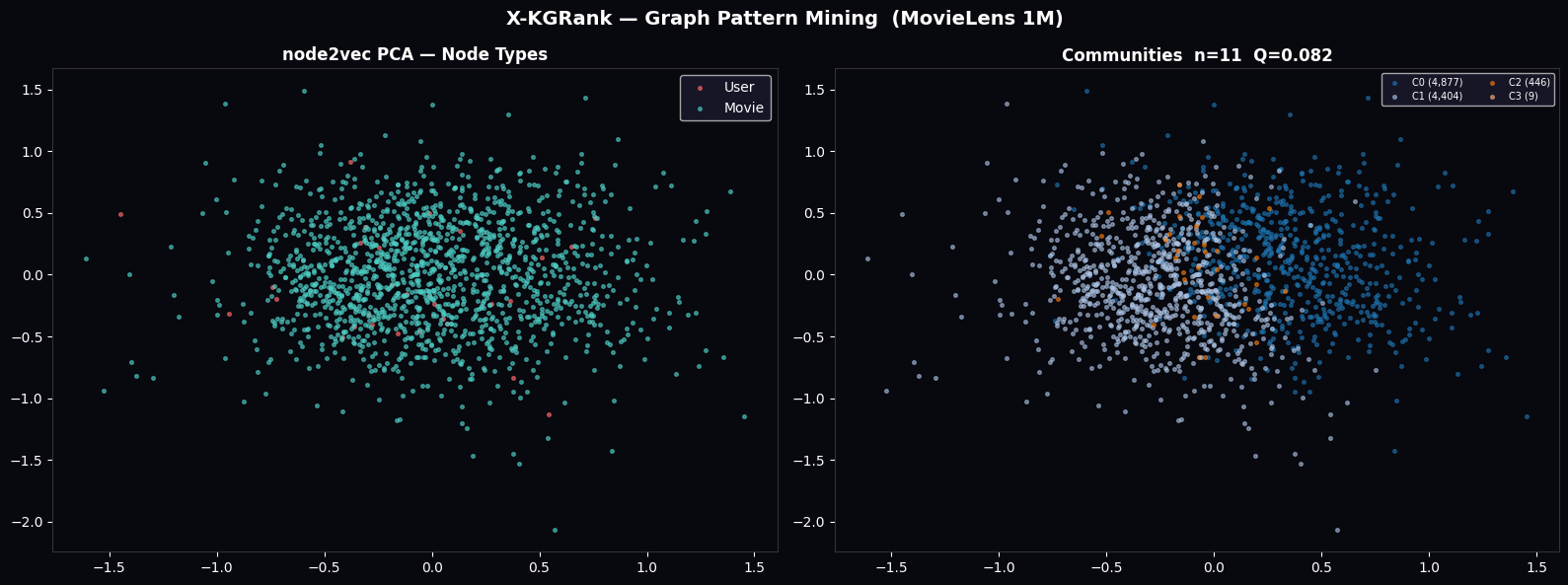}
\caption{Graph pattern mining. Left: PCA of node2vec embeddings by node type. Right: 11
communities from greedy modularity maximisation ($Q=0.082$); the two largest hold $95.1\%$
of nodes.}
\label{fig:mining}
\end{figure*}

We additionally apply greedy modularity maximisation (Clauset--Newman--Moore) and find $11$
communities with modularity $Q=0.082$. The size distribution is bimodal: two large
communities---one dominated by movies and one by users---hold $95.1\%$ of all nodes, while
the remaining nine range from $446$ nodes down to single digits. This reflects the
underlying bipartite topology, in which users and movies form the two principal components.
We do not interpret these communities as definitive categories; instead, they serve as
coarse neighbourhood context for downstream retrieval and re-ranking
(Fig.~\ref{fig:mining}).

\subsection{Popularity-Selective Routing}
For very popular items, the LLM has typically already learned the relevant facts from
pretraining; for less popular long-tail items, the knowledge graph provides concrete
behavioural evidence that the LLM lacks. Applying KG retrieval only where it is useful makes
the system efficient. We define item popularity $\pi(i)$ as the number of training
interactions involving item $i$, and the median popularity
$\pi^{*}=\mathrm{median}_{i\in\mathcal{I}}\,\pi(i)$. Items are split into two groups,
\begin{equation}
\mathcal{I}_{\text{cold}}=\{\,i:\pi(i)\le\pi^{*}\,\},
\qquad
\mathcal{I}_{\text{warm}}=\{\,i:\pi(i)>\pi^{*}\,\},
\end{equation}
where cold (long-tail) items benefit from KG retrieval and warm (head) items can be served
from pretrained knowledge without graph augmentation. The median split ($p=0.50$) yields
$1{,}855$ cold and $1{,}849$ warm items; note that interactions are not balanced, because
popular items receive many more ratings despite the two groups containing the same number of
items.

Before invoking the LLM, the system retrieves $20$ candidate items, either from an
SBERT--FAISS index for natural-language queries or from the top-scoring LightGCN candidates
for user-ID queries. It then checks each candidate's popularity. If the item is cold, the
system performs KG retrieval and extracts a $2$-hop path connecting the user to the item from
the in-memory NetworkX projection of the graph, and inserts this path into the LLM prompt as
grounding context. If the item is warm, no KG retrieval is performed and the LLM receives an
open prompt, relying on its pretrained knowledge of the item and the user's preferences. This
reduces expensive path-retrieval queries by roughly $50\%$.

\subsection{MLP Projector to the LLM Embedding Space}
The knowledge graph provides each movie with a $64$-dimensional node2vec vector, whereas the
Flan-T5-base encoder operates in a $768$-dimensional language-embedding space. To bridge this
gap we train a small neural network that maps KG vectors into the LLM space. The KG embedding
$\mathbf{z}_i^{\mathrm{KG}}\in\mathbb{R}^{64}$ captures a movie's neighbourhood, its CO\_RATED
relationships, and its position within graph communities, but is only meaningful inside the
graph embedding space; the Flan-T5-base text embedding $\mathbf{t}_i\in\mathbb{R}^{768}$
captures language meaning from the movie's title and genre. We therefore train a projector
$f_\theta:\mathbb{R}^{64}\rightarrow\mathbb{R}^{768}$, a multi-layer perceptron
\begin{equation}
f_\theta(\mathbf{z})=W_3\,\mathrm{LN}\!\big(\mathrm{GELU}(W_2\,\mathrm{LN}(\mathrm{GELU}(W_1\mathbf{z})))\big),
\end{equation}
whose output is normalised to unit length. GELU activations let the network learn complex
relationships between graph structure and language meaning, and layer normalisation stabilises
training (preventing the degenerate solution of one identical output for every movie). The
projector has $281{,}088$ trainable parameters, far fewer than the $220$M parameters of
Flan-T5-base, and aligns the two representations. For each of the $3{,}704$ movies present in
both the knowledge graph and the text metadata, we build a target text embedding by passing the
title and genre through the Flan-T5-base encoder and mean-pooling the token embeddings into a
single $768$-dimensional vector.

We use an InfoNCE objective rather than a simple reconstruction loss, because reconstruction
drives the projected vectors toward the average text vector (mean regression), whereas InfoNCE
keeps each movie individually identifiable, which is what retrieval and re-ranking require:
\begin{equation}
\mathcal{L}_{\mathrm{InfoNCE}}=
-\log\frac{\exp\!\big(\mathrm{sim}(f_\theta(\mathbf{z}_i),\mathbf{t}_i)/\tau\big)}
{\sum_{j}\exp\!\big(\mathrm{sim}(f_\theta(\mathbf{z}_i),\mathbf{t}_j)/\tau\big)},
\quad \tau=0.07 .
\end{equation}
A smaller temperature sharpens the distinction between similar and dissimilar items. The
training loss decreased from $4.01$ at epoch $10$ to $3.11$ at epoch $40$; a uniform random
projector over a batch of $256$ would yield an expected loss of $\log 256 = 5.55$, so the
achieved value represents a $44\%$ reduction, indicating that the projector has learned a
non-trivial alignment. After training, the projector is frozen so that the KG-to-LLM alignment
stays consistent. The projector establishes an alignment between the KG and LLM embedding
spaces; in the runtime pipeline reported here, however, the retrieved KG path is supplied to the
LLM as text rather than as a soft-prompt embedding, and deploying the projected embedding as a
soft prompt is left to future work.

\subsection{LLM Re-Ranking with Grounded Explanations}
A good explanation should cite real evidence: that the user liked a particular movie, that the
movie shares a genre with the recommendation, that two movies are connected by a CO\_RATED edge,
or that the candidate is linked through a graph path. Such evidence reduces hallucination.

Every recommendation should expose a path the user can understand, which makes the system more
trustworthy. The system tries to find a path between the user and the candidate, but the graph is
sometimes sparse and a clear path does not always exist. We therefore use a four-tier fallback
strategy, because the LLM always needs something to ground its prompt. (1)~\emph{Direct shortest
path}: the system finds the shortest path between the user and the movie---for example,
User~2652 rated \emph{Network} (1976) five stars---which is high-quality evidence. (2)~\emph{Shared-genre
bridge}: if no direct path is found, the system connects the user and candidate through a shared
genre; this is weaker but still understandable. (3)~\emph{CO\_RATED behavioural bridge}: if the
genre path fails, the system uses a behavioural-similarity path based on co-rating patterns rather
than content. (4)~\emph{Embedding-similarity fallback}: if all graph paths fail, the system returns
a placeholder edge---the weakest grounding, but enough to prevent complete failure.

The LLM receives a prompt containing the user profile (their five most highly rated training
items), the candidate metadata (the recommended movie and its genre), and, for cold items, the
retrieved KG path; warm items receive an open prompt. The prompt forces a fixed two-sentence
output, which keeps inference cost low and forces the model to state only the main reasons. We
compare three instruction-tuned LLMs under the same prompt: Flan-T5-large (780M parameters,
encoder--decoder), Qwen2.5-1.5B-Instruct (1.5B parameters, decoder), and Mistral-7B-Instruct-v0.2
(7B parameters, decoder). For Flan-T5 we use beam search with $4$ beams; for Qwen and Mistral we
use greedy decoding, which makes evaluation more consistent. Although the LLM produces
explanations in natural language, the final ranking is determined by LightGCN scores, so the
LLM's role is explanation over the retrieved KG path.

\section{Experimental Setup}

\subsection{Dataset}
We evaluate on MovieLens-1M, which contains roughly one million $1$--$5$ star ratings from
$6{,}040$ users on $3{,}704$ movies. The dataset is partitioned into $988{,}129$ training,
$6{,}040$ validation, and $6{,}040$ test interactions under leave-one-out splitting, with each
user contributing at least $20$ ratings ($5$-core filtering). The constructed knowledge graph
contains $9{,}762$ nodes and $999{,}264$ edges, with RATED ($988{,}129$), HAS\_GENRE ($6{,}190$),
and CO\_RATED ($4{,}945$). Table~\ref{tab:dataset} summarises the dataset statistics.

\begin{table}[t]
\caption{Dataset and Knowledge-Graph Statistics (MovieLens-1M)}
\label{tab:dataset}
\centering
\begin{tabular}{lr}
\toprule
\textbf{Statistic} & \textbf{Value}\\
\midrule
Users & 6{,}040\\
Movies & 3{,}704\\
Genres & 18\\
Train interactions & 988{,}129\\
Validation interactions & 6{,}040\\
Test interactions & 6{,}040\\
Avg.\ ratings per user & 165.4\\
KG nodes & 9{,}762\\
KG edges & 999{,}264\\
\quad RATED edges & 988{,}129\\
\quad HAS\_GENRE edges & 6{,}190\\
\quad CO\_RATED edges & 4{,}945\\
Bipartite graph density & 0.088\\
\bottomrule
\end{tabular}
\end{table}

\subsection{Evaluation Protocol}
We adopt a 99-sample protocol~\cite{krichene2020}: for each user in the test set, the held-out
positive item is paired with $99$ sampled negatives, yielding a $100$-item ranking pool. All
metrics are computed over this pool and averaged across test users.

\subsection{Metrics}
We report four standard ranking metrics: (a)~NDCG@K, which captures both relevance and position;
(b)~Recall@K, the fraction of held-out positives appearing in the top-$K$; (c)~HR@K, the fraction
of users with at least one positive in the top-$K$; and (d)~MRR, the mean reciprocal rank of the
first relevant item. We report NDCG and Recall at $K\in\{5,10,20\}$ and HR at $K=10$. For
explanation quality (Section~\ref{sec:llm}) we report a quality score that combines length
adequacy, reference rate, sentence structure, and the presence of specific reasoning keywords,
scaled to $[0,1]$. This is a proxy metric; rigorous evaluation would require human annotation.

\subsection{Baselines}
We compare against three baselines. \emph{Random} assigns each candidate in the $100$-item pool a
uniform random score. \emph{Popularity} scores each candidate by its training interaction count
$\pi(i)$; because MovieLens-1M is long-tailed, this is a strong baseline. \emph{LightGCN+SBERT} is
the structural ranker trained without KG path retrieval or LLM re-ranking, and measures the
contribution of the structural model alone.

\subsection{LLM Backbones}
The ranking and explanation stages are evaluated with three instruction-tuned LLMs:
Flan-T5-large, Qwen2.5-1.5B-Instruct, and Mistral-7B-Instruct-v0.2. The MLP projector that maps
node2vec embeddings into the LLM embedding space targets the smaller Flan-T5-base for alignment
efficiency.

\subsection{Implementation Details}
All stages are implemented in PyTorch and PyTorch Geometric, with HuggingFace Transformers for the
LLM backbones, \texttt{sentence-transformers} for SBERT encoding, NetworkX for graph construction
and traversal, and FAISS for nearest-neighbour retrieval. LightGCN training, node2vec/DeepWalk
training, and the MLP projector are trained on a single NVIDIA A100 GPU via Google Colab Pro in
fp32, and LLM inference is performed in fp16. End-to-end training across all stages takes
approximately three hours. Code, trained checkpoints, and configuration files are released at
\url{https://github.com/MeenakshiRajpurohit/graph-rag-recommend}.

\section{Results}

\subsection{Main Results}
Table~\ref{tab:main} reports ranking performance on the MovieLens-1M test set under the 99-sample
protocol.

\begin{table}[t]
\caption{Ranking Performance on MovieLens-1M (99-Sample Protocol)}
\label{tab:main}
\centering
\begin{tabular}{lccc}
\toprule
\textbf{Metric} & \textbf{Random} & \textbf{Popularity} & \textbf{X-KGRank}\\
\midrule
NDCG@5   & 0.0267 & 0.2046 & \textbf{0.2400}\\
NDCG@10  & 0.0434 & 0.2525 & \textbf{0.2956}\\
NDCG@20  & 0.0692 & 0.2983 & \textbf{0.3449}\\
Recall@5 & 0.0457 & 0.3099 & \textbf{0.3645}\\
Recall@10& 0.0980 & 0.4586 & \textbf{0.5371}\\
Recall@20& 0.2014 & 0.6401 & \textbf{0.7327}\\
HR@10    & 0.0980 & 0.4586 & \textbf{0.5371}\\
MRR      & 0.0502 & 0.2124 & \textbf{0.2435}\\
\bottomrule
\end{tabular}
\end{table}

Random ranking on a $100$-item pool with a single positive is a degenerate baseline included only
as a sanity check, so the large gap between Random and the other methods is not informative. The
meaningful comparison is X-KGRank versus Popularity, which is strong on MovieLens-1M because most
interactions concentrate on popular films; beating it requires recovering user-specific preference
signals. X-KGRank improves over Popularity by $+17.1\%$ on NDCG@10 ($0.2956$ vs.\ $0.2525$),
$+17.1\%$ on Recall@10 ($0.5371$ vs.\ $0.4586$), $+15.6\%$ on NDCG@20 ($0.3449$ vs.\ $0.2983$), and
$+14.6\%$ on MRR ($0.2435$ vs.\ $0.2124$). The relative improvement narrows at $K=20$ (Recall@20
lift is $+14.5\%$), indicating that X-KGRank surfaces additional relevant items beyond
popularity-based ranking.

\subsection{LLM Backbone Comparison}
\label{sec:llm}
To assess how the choice of language model affects explanation quality and inference cost while
holding the upstream pipeline (LightGCN, KG, KG paths, prompt) fixed, we evaluate three
instruction-tuned LLMs on $16$ test cases drawn from four randomly sampled active users. Each LLM
produces two-sentence explanations per recommendation, scored by the quality metric defined in
Section~V-C. Table~\ref{tab:llm} reports the mean quality and latency.

\begin{table}[t]
\caption{LLM Backbone Comparison (16 Cases)}
\label{tab:llm}
\centering
\begin{tabular}{lccc}
\toprule
\textbf{LLM} & \textbf{Params} & \textbf{Quality (Mean)} & \textbf{Speed (s)}\\
\midrule
Flan-T5-large            & 780M & 0.46 & 1.1\\
Qwen2.5-1.5B-Instruct    & 1.5B & 0.97 & 4.2\\
Mistral-7B-Instruct-v0.2 & 7B   & 0.94 & 4.5\\
\bottomrule
\end{tabular}
\end{table}

\begin{figure*}[!tb]
\centering
\includegraphics[width=\textwidth]{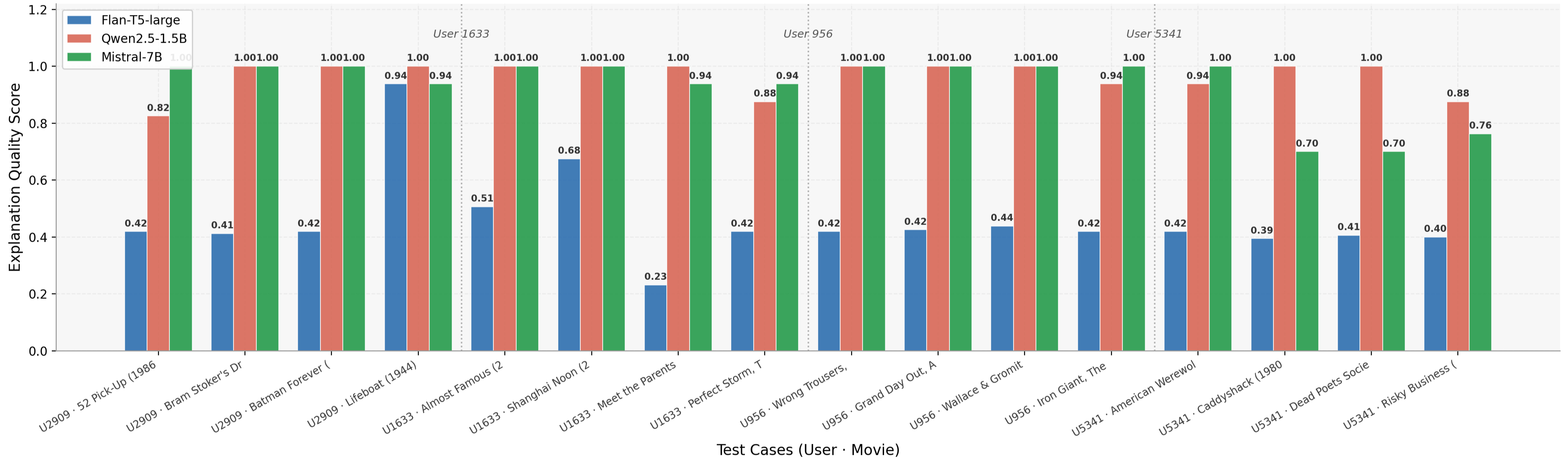}
\caption{Per-case explanation quality across 16 (user, movie) pairs for the three LLM backbones.}
\label{fig:percase}
\end{figure*}

\begin{figure*}[!tb]
\centering
\includegraphics[width=\textwidth]{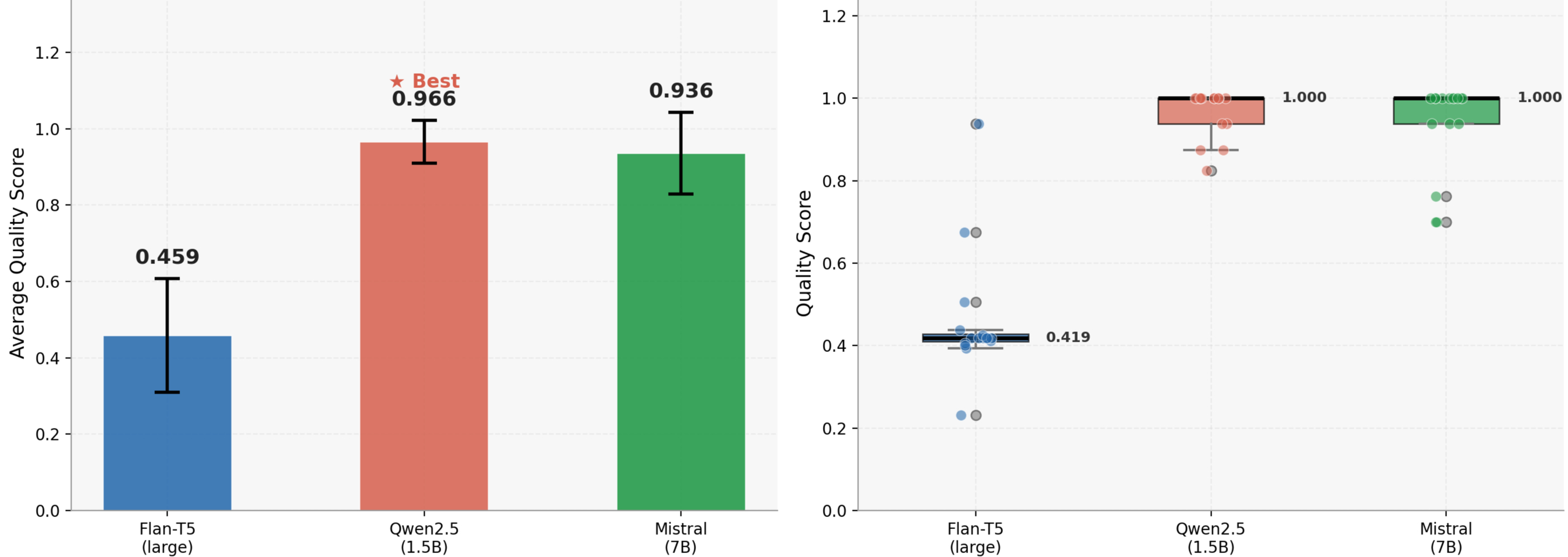}
\caption{LLM explanation quality. Left: mean $\pm$ standard deviation over 16 cases. Right: score
distributions.}
\label{fig:summary}
\end{figure*}

The per-case quality scores are reported in Fig.~\ref{fig:percase}, with the summary in
Fig.~\ref{fig:summary}. Qwen2.5-1.5B and Mistral-7B are roughly tied, and both clearly outperform
Flan-T5-large. Qwen achieves the highest mean quality ($0.97$), edging Mistral ($0.94$) by a small
margin. The gap between the two is dominated by a handful of cases in which Mistral produces
tokeniser artefacts (e.g., concatenated tokens such as ``amust watch'' or ``BraveHeart(19five)'')
that lower its score; these are surface-level glitches rather than substantive content failures.
Flan-T5-large lags far behind at $0.46$, producing truncated, repetitive, and partial explanations
across many cases; its only advantage is speed ($1.1$\,s vs.\ $4.2$ and $4.5$\,s).

That a $1.5$B-parameter model matches a $7$B-parameter model on this task is the most substantive
efficiency finding from these comparisons: Qwen2.5-1.5B achieves marginally higher explanation
quality than Mistral-7B at less than a quarter of the parameter count and identical latency. Where
memory and cost are constraints, Qwen offers a clear benefit. We interpret this as evidence that
grounded explanation generation---where the KG path is retrieved---reduces dependence on model
scale, since much of the factual content is supplied externally by the KG path rather than recalled
from the model's parameters.

The quality metric, however, is a proxy. It measures properties such as sentence length, title
reference, sentence structure, and keyword presence, but does not directly assess factual accuracy.
Manual inspection of high-scoring outputs reveals factual errors, including misattributed directors,
fabricated co-stars, and incorrect release years (documented in Section~\ref{sec:qual}). The
$16$-case comparison is sufficient to rank the three models (Qwen $\approx$ Mistral $\gg$ Flan-T5)
but does not support the precise Qwen-versus-Mistral margin.

\section{Qualitative Analysis}
\label{sec:qual}

To illustrate X-KGRank's behaviour concretely, we walk through cases drawn from a
natural-language RAG query demonstration. The first shows the system working as intended; the
second shows genuine limitations. Both use the SBERT--FAISS configuration for candidate retrieval
and KG path extraction.

\subsection{Success Case: ``Moving drama about family and loss''}
For this query, SBERT--FAISS retrieved six candidates; the top three were \emph{My Family} (1995,
score $0.60$), \emph{Two Family House} (2000, score $0.56$), and \emph{The Funeral} (1996, score
$0.54$). All three are dramas aligned with the query, indicating that the SBERT semantic match
operates correctly. For each candidate, the system extracted a two-hop KG path connecting the user
to the candidate through a shared dramatic film:

{\footnotesize\ttfamily
\#1 My Family (1995)\\
\indent User 1 --[RATED 4.0]--> Sixth Sense, The (1999)\\
\indent Sixth Sense, The (1999) --[RATED 5.0]--> User 62\\[2pt]
\#2 Two Family House (2000)\\
\indent User 1 --[RATED 4.0]--> E.T. the Extra-Terrestrial (1982)\\
\indent E.T. the Extra-Terrestrial (1982) --[RATED 4.0]--> User 173\\[2pt]
\#3 The Funeral (1996)\\
\indent User 1 --[RATED 5.0]--> Schindler's List (1993)\\
\indent Schindler's List (1993) --[RATED 3.0]--> User 225\\
}

\noindent These paths surface meaningful behaviour: viewers who rated films like \emph{Sixth Sense},
\emph{E.T.}, and \emph{Schindler's List} also engage with highly dramatic family films. The
intermediate users (62, 173, 225) are not semantically meaningful in themselves; they are the
structure through which the recommendations acquire behavioural grounding. The Flan-T5-large
explanations are weak in this case, restating the query rather than reasoning from the retrieved
path.

\subsection{Failure Case: ``Mind-bending thriller with unexpected plot twist''}
\label{sec:failure}
For this query, SBERT--FAISS retrieved \emph{What Lies Beneath} (2000, similarity $0.55$),
\emph{A Simple Plan} (1998, $0.54$), and \emph{Twister} (1996, $0.54$). The first two are
appropriate thrillers that match the query intent. \emph{Twister}, however, is a tornado-disaster
film whose genre includes ``Thriller'' but whose narrative---storm-chasing scientists pursuing
tornadoes---has no relationship to a mind-bending or twist-driven plot. This is a clear retrieval
error. The retrieved KG paths also reveal a limitation:

{\footnotesize\ttfamily
\#1 What Lies Beneath (2000)\\
\indent User 1 --[RATED 4.0]--> Girl, Interrupted (1999)\\
\indent Girl, Interrupted (1999) --[RATED 3.0]--> User 90\\[2pt]
\#3 Twister (1996)\\
\indent User 1 --[RATED 4.0]--> Girl, Interrupted (1999)\\
\indent Girl, Interrupted (1999) --[RATED 3.0]--> User 90\\
}

\noindent The paths for \emph{What Lies Beneath} and \emph{Twister} are identical: both route through
the same intermediate film (\emph{Girl, Interrupted}) and the same intermediate user (90). Because
the demo user has a limited rating history, the shortest-path search repeatedly returns the same
high-degree bridge node regardless of the candidate, producing paths that carry no
candidate-specific information; the KG path therefore does not discriminate between the appropriate
\emph{What Lies Beneath} and the inappropriate \emph{Twister}. The LLM compounds both errors: for
\emph{Twister}, Flan-T5-large generates ``\emph{Twister} is a mind-bending thriller with unexpected
plot twists,'' a factually false assertion that simply projects the query words onto the candidate
rather than reasoning from evidence. This case exposes two limitations: retrieval operates on genre
tags rather than narrative content, so any film tagged ``Thriller'' is eligible for thriller-intent
queries; and KG paths degrade when users with sparse history fail to differentiate candidates.

\subsection{LLM Case Study: Factual Accuracy versus Fluency}
To illustrate the qualitative differences identified in Section~\ref{sec:llm}, we examine the
explanations generated by Qwen2.5-1.5B and Mistral-7B for an identical set of recommendations.
User~2909's profile spans classic and mid-century cinema---\emph{Casablanca} (1942),
\emph{Double Indemnity} (1944), \emph{Lawrence of Arabia} (1962), \emph{Back to the Future} (1985),
and \emph{Murder in the First} (1995)---and LightGCN ranked four candidates reached by distinct KG
paths:

{\footnotesize\ttfamily
52 Pick-Up (1986):\\
\indent User 2909 --[RATED 4.0]--> My Fair Lady --[RATED 4.0]--> User 183\\[2pt]
Bram Stoker's Dracula (1992):\\
\indent User 2909 --[RATED 4.0]--> Awakenings --[CO\_RATED w=198]--> Dracula\\[2pt]
Batman Forever (1995):\\
\indent User 2909 --[RATED 5.0]--> Payback --[CO\_RATED w=263]--> Batman Forever\\[2pt]
Lifeboat (1944):\\
\indent User 2909 --[RATED 4.0]--> North by Northwest --[RATED 3.0]--> User 23\\
}

\noindent Here \emph{Bram Stoker's Dracula} and \emph{Batman Forever} are reached through CO\_RATED
edges, while \emph{52 Pick-Up} and \emph{Lifeboat} are reached through shared-user RATED bridges.

Mistral-7B produces factually accurate explanations. For \emph{Batman Forever} it correctly
identifies the cast---``an iconic performance by Val Kilmer as Batman and Jim Carrey as The
Riddler''---and for \emph{Lifeboat} it gives an accurate plot summary: ``a group of strangers are
stranded at sea on a lifeboat after their ship is sunk by a German U-boat.'' These are correct,
verified, and specific. Mistral's weakness is tokenisation defects such as the concatenated
``amust watch.''

Qwen2.5-1.5B produces fluent but fabricated explanations. For \emph{52 Pick-Up}, Qwen confidently
states that the film was ``directed and co-written by John Landis, with an ensemble cast including
Dan Aykroyd as Frank McHale Jr.''---a fabrication. The 1986 film was directed by John Frankenheimer
from an Elmore Leonard novel and features neither Landis nor Aykroyd, and Qwen additionally
mistakes the year as 1973. The output is grammatically polished and rich in specific-sounding
detail, which is precisely what makes the fabrication dangerous: the heuristic quality metric ranks
this fluent-but-false output above Mistral's accurate one.

\section{Limitations}

We analyse the limitations of X-KGRank to contextualise its contributions and guide future work.
\emph{(a)~Single domain.} All results are obtained on MovieLens-1M; we have not validated the
framework on another domain such as e-commerce or news. \emph{(b)~Genre-tag retrieval.} As the
failure case in Section~\ref{sec:failure} shows, SBERT--FAISS retrieves over movie titles and genre tags
rather than plot content; richer content representations (plot summaries, reviews) would mitigate
this but are left to future work. \emph{(c)~Path degeneracy for sparse histories.} When a user has
few highly rated films, the shortest-path search returns the same high-degree bridge node,
producing candidate-independent paths. \emph{(d)~Heuristic explanation metric.} Our quality metric
captures sentence length, title reference, structure, and keyword presence, but does not assess
factual accuracy; manual inspection confirms that even high-scoring explanations contain factual
errors, and rigorous evaluation would require human annotation. \emph{(e)~Low community modularity.}
The node2vec communities have modularity $Q=0.082$, which is low for well-separated networks; we use
them only as weak contextual features and do not rely on hard partitioning, but more structured
graphs or alternative community methods are left to future exploration.

\section{Conclusion}

We presented X-KGRank, a knowledge-graph retrieval-augmented framework for explainable
recommendation that integrates structural collaborative filtering, graph pattern mining, and
LLM-based explanation generation. We constructed a knowledge graph from MovieLens-1M and trained a
LightGCN model with SBERT-based content-aware initialisation and a rating-weighted BPR objective. A
popularity-selective routing strategy applies expensive KG retrieval only to long-tail items. On
MovieLens-1M, the full pipeline improved over the popularity baseline by $17\%$ on NDCG@10 and
Recall@10. A comparison across three LLMs shows that Qwen2.5-1.5B matches the $7$B-parameter
Mistral-7B on explanation quality, yet qualitative analysis shows that smaller models are more prone
to factual errors---suggesting that external KG grounding reduces the dependence of recommendation
quality on model scale, while factual explanation reliability still benefits from larger LLMs.
Future directions include human evaluation of explanation quality, validation across additional
domains, and content-attribute knowledge graphs built from plot summaries and reviews.

\section{Project Artifacts and Implementation Resources}

Because this project involved a complete implementation of the X-KGRank
framework, all source code, notebooks, datasets, and visualisations have
been organised and made publicly accessible. These resources cover the full
pipeline---from knowledge-graph construction and structural representation
learning to popularity-selective routing, KG path retrieval, and LLM-based
explanation---ensuring transparency, reproducibility, and ease of
verification.

\subsection{Source Code Repository}
The complete Python implementation---including the knowledge-graph
construction scripts, the LightGCN ranker, graph pattern mining, the MLP
projector, and the LLM explanation pipeline---is hosted on GitHub:
\begin{center}
\url{https://github.com/MeenakshiRajpurohit/graph-rag-recommend}
\end{center}

\balance


\begin{thebibliography}{99}

\bibitem{lightgcn}
X.~He, K.~Deng, X.~Wang, Y.~Li, Y.~Zhang, and M.~Wang,
``LightGCN: Simplifying and Powering Graph Convolution Network for Recommendation,''
in \emph{Proc.\ 43rd Int.\ ACM SIGIR Conf.\ Research and Development in Information Retrieval},
2020, pp.~639--648.

\bibitem{graphsage}
W.~L. Hamilton, R.~Ying, and J.~Leskovec,
``Inductive Representation Learning on Large Graphs,''
in \emph{Advances in Neural Information Processing Systems (NeurIPS)}, 2017, pp.~1024--1034.

\bibitem{kgat}
X.~Wang, X.~He, Y.~Cao, M.~Liu, and T.-S. Chua,
``KGAT: Knowledge Graph Attention Network for Recommendation,''
in \emph{Proc.\ 25th ACM SIGKDD Int.\ Conf.\ Knowledge Discovery \& Data Mining}, 2019, pp.~950--958.

\bibitem{p5}
S.~Geng, S.~Liu, Z.~Fu, Y.~Ge, and Y.~Zhang,
``Recommendation as Language Processing (RLP): A Unified Pretrain, Personalized Prompt \& Predict Paradigm (P5),''
in \emph{Proc.\ 16th ACM Conf.\ Recommender Systems (RecSys)}, 2022, pp.~299--315.

\bibitem{hou2024}
Y.~Hou, J.~Zhang, Z.~Lin, H.~Lu, R.~Xie, J.~McAuley, and W.~X. Zhao,
``Large Language Models are Zero-Shot Rankers for Recommender Systems,''
in \emph{Proc.\ European Conf.\ Information Retrieval (ECIR)}, 2024.

\bibitem{tallrec}
K.~Bao, J.~Zhang, Y.~Zhang, W.~Wang, F.~Feng, and X.~He,
``TALLRec: An Effective and Efficient Tuning Framework to Align Large Language Model with Recommendation,''
in \emph{Proc.\ 17th ACM Conf.\ Recommender Systems (RecSys)}, 2023, pp.~1007--1014.

\bibitem{gretriever}
X.~He, Y.~Tian, Y.~Sun, N.~V. Chawla, T.~Laurent, Y.~LeCun, X.~Bresson, and B.~Hooi,
``G-Retriever: Retrieval-Augmented Generation for Textual Graph Understanding and Question Answering,''
in \emph{Advances in Neural Information Processing Systems (NeurIPS)}, 2024.

\bibitem{kragrec}
S.~Wang et al.,
``K-RagRec: Knowledge Graph Retrieval-Augmented Generation for LLM-based Recommendation,''
\emph{arXiv preprint} arXiv:2501.02226, 2025.

\bibitem{krichene2020}
W.~Krichene and S.~Rendle,
``On Sampled Metrics for Item Recommendation,''
in \emph{Proc.\ 26th ACM SIGKDD Int.\ Conf.\ Knowledge Discovery \& Data Mining}, 2020, pp.~1748--1757.

\end{thebibliography}
\end{document}